\documentclass[twocolumn,floatfix,showpacs,prb]{revtex4}
\usepackage{graphicx}
\usepackage{amsmath}
\usepackage{amssymb}
\usepackage{bm}

\begin{document}

\title{Random-matrix theory of thermal conduction in superconducting quantum dots}
\author{J. P. Dahlhaus, B. B\'{e}ri, and C. W. J. Beenakker}
\affiliation{Instituut-Lorentz, Universiteit Leiden, P.O. Box 9506, 2300 RA Leiden, The Netherlands}
\date{April 2010}
\begin{abstract}
We calculate the probability distribution of the transmission eigenvalues $T_{n}$ of Bogoliubov quasiparticles at the Fermi level in an ensemble of chaotic Andreev quantum dots. The four Altland-Zirnbauer symmetry classes (determined by the presence or absence of time-reversal and spin-rotation symmetry) give rise to four circular ensembles of scattering matrices. We determine $P(\{T_{n}\})$ for each ensemble, characterized by two symmetry indices $\beta$ and $\gamma$. For a single $d$-fold degenerate transmission channel we thus obtain the distribution $P(g)\propto g^{-1+\beta/2}(1-g)^{\gamma/2}$ of the thermal conductance $g$ (in units of $d\pi^{2}k_{B}^{2}T_{0}/6h$ at low temperatures $T_{0}$). We show how this single-channel limit can be reached using a topological insulator or superconductor, without running into the problem of fermion doubling. 
\end{abstract}
\maketitle

\section{Introduction}
\label{intro}

The Landauer approach to quantum transport\cite{Dat95,Hou96,Imr99} relates a transport property (such as the electrical or thermal conductance) to the eigenvalues $T_{n}$ of the transmission matrix product $tt^\dagger$. If transport takes place through a region with chaotic scattering (typically a quantum dot), random-matrix theory (RMT) provides a statistical description.\cite{Bee97,Guh98,Alh00} While the properties of individual chaotic systems are highly sensitive to the microscopic parameters of the scattering region, such as its geometry or the arrangements of impurities, they obey universal statistical features, independent of  these details, on energy scales below the Thouless energy (the inverse of the dwell time). The distribution $P(\{T_n\})$ of the transmission eigenvalues then naturally emerges as the determining quantity for the distribution of the transport properties.

While microscopic details do not influence the statistics, the role of symmetries is essential.  According to Dyson,\cite{Dys62,Meh91} there are three symmetry classes in normal (non-superconducting) electronic systems, characterized by a symmetry index $\beta$ depending on the presence or absence of time-reversal and spin-rotation symmetry (cf.\ Table \ref{tab:table1}). The transmission eigenvalue distribution for these three RMT ensembles is known.\cite{Bar94,Jal94} For a single $d$-fold degenerate channel at the entrance and exit of the quantum dot this gives the distribution
\begin{equation}
P(g)\propto g^{-1+\beta/2},\;\;0<g<1,\label{PTDyson}
\end{equation}
of the electrical conductance $g$ (in units of $de^{2}/h$).
The full distribution $P(\{T_{n}\})$ has found a variety of physical applications,\cite{Bee09} and has also been used in a more mathematical context to obtain exact results for electrical conductance and shot noise\cite{Sav06,Kho09} and to uncover connections between quantum chaos and integrable models.\cite{Osi08} 

As first shown by Altland and Zirnbauer,\cite{Alt97} Dyson's classification scheme becomes insufficient in the presence of superconducting order: The particle-hole symmetry of the Bogoliubov-De Gennes Hamiltonian produces four new symmetry classes.\cite{Hei05,Ryu09,Alt10} Depending again on the presence or absence of time-reversal and spin-rotation symmetry, these classes are characterized by $\beta$ and a second symmetry index $\gamma$ (cf.\ Table \ref{tab:table2}).\cite{Cas96,note2} As we show in this paper, the analogous result to Eq.\ \eqref{PTDyson} is
\begin{equation}
P(g)\propto g^{-1+\beta/2}(1-g)^{\gamma/2},\;\;0<g<1,\label{PTAZ}
\end{equation}
where now $g$ is the \textit{thermal} conductance in units of $d\pi^{2}k_{B}^{2}T_{0}/6h$ (at temperature $T_{0}$). We consider thermal transport instead of electrical transport because the Bogoliubov quasiparticles that are transmitted through a superconducting quantum dot carry a definite amount of energy rather than a definite amount of charge. (Charge is not conserved upon Andreev reflection at the superconductor, when charge-$2e$ Cooper pairs are absorbed by the superconducting condensate.)

\begin{table*}
\centering
\begin{tabular}{ | l || c | c | c |}
\hline
Ensemble name & \textbf{CUE} & \textbf{COE} & \textbf{CSE} \\ \hline
Symmetry class\ &  A &  AI & AII \\ \hline
$S$-matrix elements\ &  complex &  complex & complex \\ \hline
$S$-matrix space & unitary & unitary symmetric & unitary selfdual\\ \hline \hline
Time-reversal symmetry & $\times$ & $S=S^{T}$ &  $S=\sigma_{2}S^{T}\sigma_{2}$ \\ \hline
Spin-rotation symmetry &  $\times$ or \checkmark &  \checkmark & $\times$ \\ \hline \hline
degeneracy $d$ of $T_{n}$  &  $1$ or $2$ &  $2$ & $2$ \\ \hline
\qquad\qquad $\beta$ &$2$ &$1$& $4$\\ \hline
\end{tabular}
\caption{Classification of the Wigner-Dyson scattering matrix ensembles for normal (non-superconducting) systems, with the parameter $\beta$ in the distribution \eqref{PTDyson} of the electrical conductance. (The parameter $\gamma\equiv 0$ in these ensembles.) The abbreviations C(U,O,S)E signify Circular (Unitary,Orthogonal,Symplectic) Ensemble. The Pauli matrix $\sigma_{j}$ acts on the spin degree of freedom.}
\label{tab:table1}
\end{table*}

\begin{table*}
\centering
\begin{tabular}{ | l || c | c | c | c |}
\hline
Ensemble name & \textbf{CRE} & \textbf{T-CRE} & \textbf{CQE} & \textbf{T-CQE} \\ \hline
Symmetry class\ &  D &  DIII & C & CI \\ \hline
$S$-matrix elements\ &  real &  real & quaternion & quaternion \\ \hline
$S$-matrix space & orthogonal & orthogonal selfdual & symplectic & symplectic symmetric \\ \hline \hline
Particle-hole symmetry & $S=S^{\ast}$ & $S=S^{\ast}$ & $S=\tau_{2}S^{\ast}\tau_{2}$ & $S=\tau_{2}S^{\ast}\tau_{2}$ \\ \hline
Time-reversal symmetry & $\times$ & $S=\sigma_{2}S^{T}\sigma_{2}$ &  $\times$ & $S=S^{T}$ \\ \hline
Spin-rotation symmetry &  $\times$ &  $\times$ & \checkmark & \checkmark \\ \hline \hline
degeneracy $d$ of $T_{n}$  &  $1$ & $2$ & $4$ & $4$ \\ \hline
\qquad\qquad $\beta$ &$1$&$2$&$4$&$2$\\ \hline
\qquad\qquad $\gamma$&$-1$ &$-1$& $2$&$1$\\ \hline
\end{tabular}
\caption{Classification of the Altland-Zirnbauer scattering matrix ensembles for superconducting systems. For each ensemble the parameters $\beta,\gamma$ in the distribution \eqref{PTAZ} of the thermal conductance are indicated. The Pauli matrices $\sigma_{j}$ and $\tau_{j}$ act on, respectively the spin and particle-hole degrees of freedom. The abbreviations (T)-C(R,Q)E signify (Time-reversal-symmetric)-Circular (Real,Quaternion) Ensemble.}
\label{tab:table2}
\end{table*}

Concerning previous related studies, we note that the electrical conductance has been investigated by Altland and Zirnbauer,\cite{Alt97} but not the thermal conductance. Thermal transport in superconductors has been studied in connection with the thermal quantum Hall effect in two dimensions,\cite{Sen00,Vis01,Cha02} and also in connection with one-dimensional localization.\cite{Bro00,Tit01} The present study complements these works by addressing the zero-dimensional regime in connection with chaotic scattering.   

The outline of this paper is as follows. Sections \ref{formulation} and \ref{sec:pdf} formulate the problem and present $P(\{T_{n}\})$. In Sec.\ \ref{secPG} we then apply this to the statistics of the thermal conductance. The probability distribution \eqref{PTAZ} in the single-channel limit is of particular interest (since it is furthest from a Gaussian), but it can only be reached in the Andreev quantum dot in the presence of spin-rotation symmetry. A fermion-doubling problem stands as an obstacle when spin-rotation symmetry is broken. We show how to overcome this obstacle in Sec.\ \ref{sec:topo} using topological phases of matter \cite{Vol03,Qi10,Has10} (topological superconductors or insulators). We close in Sec.\ \ref{conclusion} with a summary and a proposal to realize the superconducting ensembles in graphene.

\section{Formulation of the problem}
\label{formulation}

\subsection{Andreev quantum dot}
\label{sec:AQD}

\begin{figure}[tb]
\centerline{\includegraphics[width=0.6\linewidth]{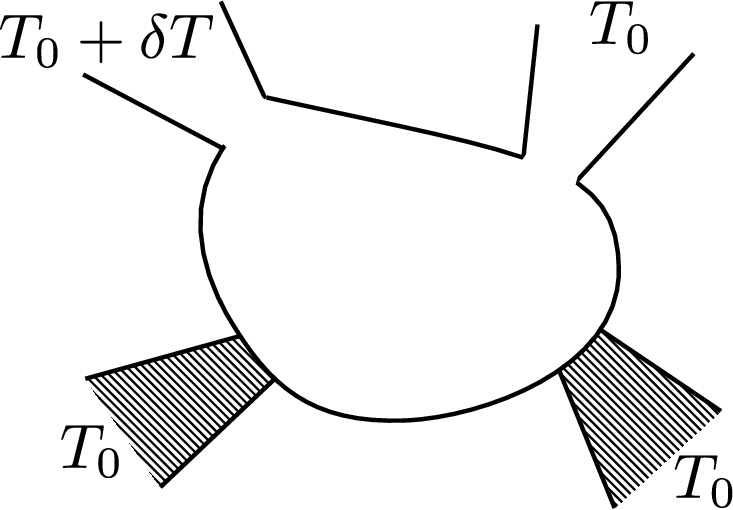}}
\caption{\label{fig:AQD}
Quantum dot in a two-dimensional electron gas, connected to a pair of superconductors (shaded) and to two normal-metal reservoirs. One of the normal reservoirs is at a slightly elevated temperature $T_{0}+\delta T$.
}
\end{figure}

An Andreev quantum dot, or Andreev billiard, is a confined region in a two-dimensional electron gas connected to superconducting electrodes (see Fig.\ \ref{fig:AQD}). Electronic transport through this system is governed by the interplay of chaotic scattering at the boundaries of the quantum dot and Andreev reflection at the superconductors. (See Ref.\ \onlinecite{Bee05} for a review.) We assume $s$-wave superconductors, with an isotropic gap $\Delta$, so for excitation energies $E<\Delta$ there are no modes propagating into the superconductors. In order to enable quasiparticle transport, the cavity has two additional leads connected to it which support $N_{1},N_{2}$ propagating modes (not counting degeneracies). The leads connect the cavity to normal-metal reservoirs in local thermal equilibrium.

Quasiparticle transmission is possible only if the excitations of the Andreev quantum dot (without the leads) are gapless. This is also necessary for the excitations to explore the phase space of the cavity, an essential requirement for chaotic scattering. Gapless excitations are ensured by taking two superconducting electrodes with the same contact resistance and a phase difference $\pi$. This value of the phase difference closes the gap while respecting time-reversal invariance (because phase differences $\pi$ and $-\pi$ are equivalent). Time-reversal invariance can be broken by application of a magnetic field, perpendicular to the plane of the dot. (A sufficiently strong magnetic field closes the gap, so then the $\pi$-phase difference of the superconductors is not needed and a single superconducting electrode is sufficient.) Spin-rotation symmetry can be broken by spin-orbit coupling. An ensemble of chaotic systems can be generated, for example, by varying the shape of the quantum dot or by a random arrangement of impurities.

In global equilibrium the superconducting and normal-metal contacts are all at the same temperature $T_{0}$ and Fermi energy (or chemical potential) $E_{F}$. For thermal conduction in the linear response regime we raise the temperature of one of the normal metals by an amount $\delta T\ll T_{0}$. The thermal conductance $G$ is the heat current between the normal reservoirs divided by $\delta T$. (The reservoirs are kept at the same chemical potential, so there is no thermo-electric contribution to the heat current.) 

If $k_{B}T_{0}$ is small compared to the Thouless energy (the inverse dwell time in the quantum dot), then $G$ is determined by the transmission eigenvalues at the Fermi energy,
\begin{equation}
G=dG_{0}\sum_{n}T_{n}.\label{Gdef}
\end{equation}
The sum runs over the ${\rm min}\,(N_{1},N_{2})$ nonzero transmission eigenvalues $T_{n}$, with spin and/or particle-hole degeneracy accounted for by the factor $d$. The thermal conductance quantum for superconducting systems is $G_{0}=\pi^{2}k_{B}^{2}T_{0}/6h$, one-half the normal-state value.\cite{Hou96,Siv86}

\subsection{Scattering matrix ensembles}
\label{Smatrixens}

The scattering matrix $S$ is a unitary matrix of dimension $(N_{1}+N_{2})\times(N_{1}+N_{2})$ that relates the amplitudes of outgoing and incoming modes in the two leads connected to the normal reservoirs. The energy is fixed at the Fermi level ($E=0$). Four sub-blocks of $S$ define the transmission and reflection matrices,
\begin{equation}
S=
\begin{pmatrix}
r_{N_{1}\times N_{1}} & t'_{N_{1}\times N_{2}}\\
t_{N_{2}\times N_{1}} & r'_{N_{2}\times N_{2}}
\end{pmatrix}.
\label{SMatrix}
\end{equation}
(The subscripts refer to the dimension of the blocks.) Table \ref{tab:table2} lists the Altland-Zirnbauer symmetry classes to which $S$ belongs, and the corresponding RMT ensembles.\cite{Alt97,Hei05,Ryu09,Alt10} We briefly discuss the various entries in that table.

In the case of systems without spin-rotation symmetry, it is convenient to choose the Majorana basis in which $S$ has real matrix elements.\cite{note1} Without time-reversal symmetry (symmetry class D), the scattering matrix space is thus the orthogonal group. The presence of time-reversal symmetry imposes the additional constraint $S=\sigma_{2} S^{T} \sigma_{2}$, where $\sigma_{j}$ is a Pauli matrix in spin-space, and $T$ indicates the matrix transpose. The scattering matrices in this symmetry class DIII are self-dual orthogonal matrices. (The combination $\sigma_{2} A^{T} \sigma_{2}$ is the so-called dual of the matrix $A$.)

If spin-rotation symmetry is preserved, the spin degree of freedom can be omitted if we use the electron-hole basis (rather than the Majorana basis). The electron-hole symmetry relation then reads $S=\tau_{2} S^{*} \tau_{2}$, where now the Pauli matrices $\tau_{j}$ act on the electron-hole degree of freedom. The matrix elements of $S$ can be written in the quaternion form $a_{0}\tau_{0}+i\sum_{n=1}^{3}a_{n}\tau_{n}$, with real coefficients $a_{n}$. The scattering matrix space for the symmetry class C without time-reversal symmetry is the symplectic group, additionally restricted to symmetric matrices in the presence of time-reversal symmetry (class CI).  

Henceforth we assume that the quantum dot is connected to the leads via ballistic point contacts. The RMT ensembles in this case are defined by $S$ being uniformly distributed with respect to the invariant measure $d\mu(S)$ in the scattering matrix space for each particular symmetry class.\cite{Alt97} (For the distribution in the case that the contacts contain tunnel barriers, see Ref.\ \onlinecite{Ber09}.) 

It is convenient to have names for the Altland-Zirnbauer ensembles, analogous to the existing names for the Dyson ensembles. Zirnbauer\cite{Alt10} has stressed that the names D,DIII,C,CI given to the symmetry classes (derived from Cartan's classification of symmetric spaces) should be kept distinct from the ensembles, because a single symmetry class can produce different ensembles. Following Ref.\ \onlinecite{Ser10}, we will refer to the Circular Real Ensemble (CRE) and Circular Quaternion Ensemble (CQE) of uniformly distributed real or quaternion unitary matrices. The presence of time-reversal symmetry is indicated by T-CRE and T-CQE. (The prefix T can also be thought of as referring to the matrix transpose in the restrictions imposed by time-reversal symmetry.)

\section{Transmission eigenvalue distribution}
\label{sec:pdf}

\subsection{Joint probability distribution}
\label{joint}

Because of unitarity, the matrix products $tt^{\dagger}$ and $t'{t'}^{\dagger}$ have the same set $T_{1},T_{2},\ldots T_{N_{\rm min}}$ of nonzero eigenvalues, with $N_{\rm min}={\rm min}\,(N_{1},N_{2})$. The calculation of the joint probability distribution $P(\{T_n\})$ of these transmission eigenvalues from the invariant measure $d\mu(S)$ is outlined in App.\ \ref{calculationPT}.\cite{For10} (It is equivalent to the calculation of the Jacobian given in Ref.\ \onlinecite{Bro00}.) The result is
\begin{align}
P(\{T_{n}\}) \propto{}& \prod_{i} T_{i}^{(\beta/2)|N_1-N_2|}T_{i}^{-1+\beta/2}(1-T_{i})^{\gamma/2}\nonumber\\
& \times\prod_{j<k} \bigl|T_{k}-T_{j} \bigr|^{\beta}. \label{pdf}
\end{align}
The values of the parameters $\beta$ and $\gamma$ characterizing the Altland-Zirnbauer symmetry classes are listed in Table \ref{tab:table2}. 

\begin{figure}[tb]
\centerline{\includegraphics[width=0.8\linewidth]{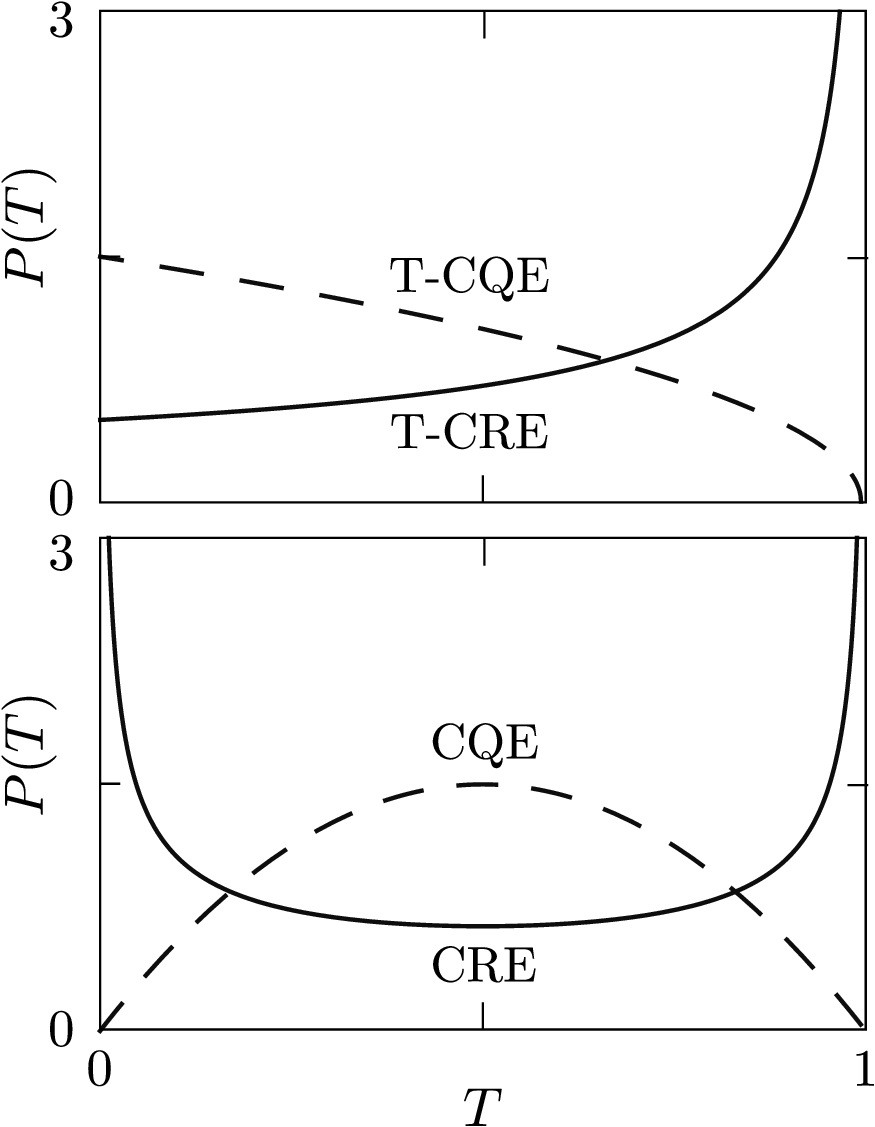}}
\caption{\label{fig:pdfPlot}
Probability distribution \eqref{pdf} in the case $N_{1}=N_{2}=1$ of a single ($d$-fold degenerate) transmission eigenvalue $T$, which then corresponds to the (dimensionless) thermal conductance $g=G/dG_{0}$. The four curves correspond to the four superconducting ensembles in Table \ref{tab:table2}. 
}
\end{figure}

The distribution \eqref{pdf} differs from the result\cite{Bee97,Bar94,Jal94} in the Dyson ensembles by the factor $\prod_{i} (1-T_{i})^{\gamma/2}$. Depending on the sign of $\gamma$, this factor produces a repulsion or attraction of the $T_{i}$'s to perfect \textit{transmission}. In contrast, the factor $\prod_{i}T_{i}^{-1+\beta/2}$, which exists also in the Dyson ensembles, repels or attracts the $T_{i}$'s to perfect \textit{reflection}. The distributions $P(T_{1})$ for $N_{1}=N_{2}=1$ in the various ensembles are plotted in Fig.\ \ref{fig:pdfPlot}. In view of Eq.\ \eqref{Gdef}, this is just the distribution \eqref{PTAZ} of the thermal conductance in the single-channel limit announced in the Introduction. (How to actually reach this limit is discussed in following Sections.)

\subsection{Eigenvalue density}
\label{sec_rhoT}

\begin{figure}[tb]
\centerline{\includegraphics[width=0.9\linewidth]{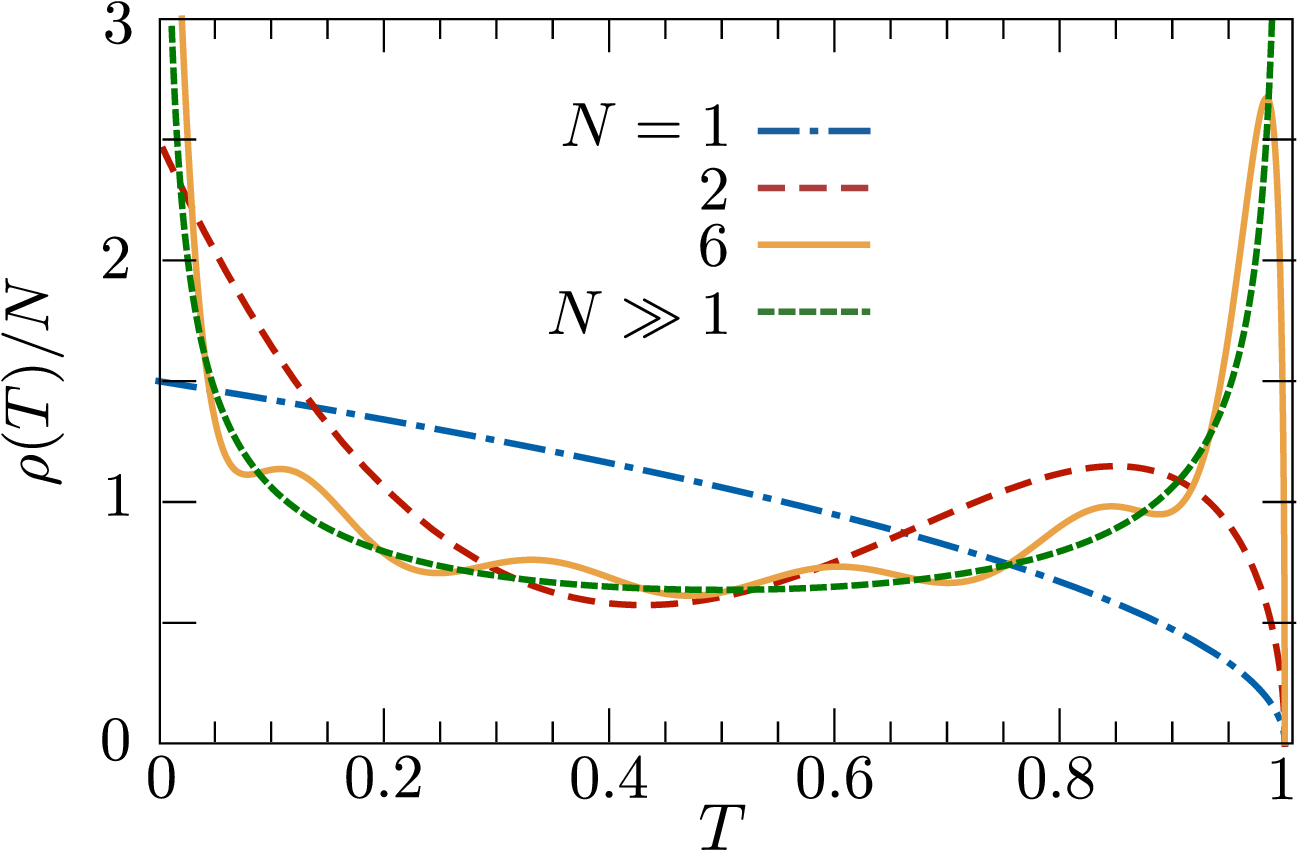}}
\caption{\label{fig:rhoT}
Transmission eigenvalue densities in the T-CQE for various numbers $N=N_{1}=N_{2}$ of transmission eigenvalues, calculated from Eq.\ \eqref{pdf}. The large-$N$ limit is the same for each ensemble.
} 
\end{figure}

The density $\rho(T)$ of the transmission eigenvalues is defined by
\begin{equation}
\rho(T)=\left\langle\sum_{n}\delta(T-T_{n})\right\rangle,\label{rhoTdef}
\end{equation}
where $\langle\cdots\rangle$ denotes an average with distribution \eqref{pdf}. It can be calculated for $N_{1}, N_{2}\gg 1$ using the general methods of RMT.\cite{Bee97}

To leading order in $N_{1},N_{2}$ the eigenvalue density approaches the $\beta$ and $\gamma$ independent limiting form\cite{Bee97,Bar94,Jal94}
\begin{align}
\rho_{0}(T)={}&\frac{N_{1}+N_{2}}{2\pi}\left(\frac{T-T_{c}}{1-T}\right)^{1/2}\frac{1}{T}\nonumber\\
&\times\Theta(1-T)\Theta(T-T_{c}), \label{rho0}
\\
T_{c}={}&\frac{(N_{1}-N_{2})^{2}}{(N_{1}+N_{2})^{2}}.\label{Tcdef}
\end{align}
(The function $\Theta(x)$ is the unit step function, $\Theta(x)=0$ if $x<0$ and $\Theta(x)=1$ if $x>0$.) The approach to this ensemble-independent density with increasing $N_{1}=N_{2}$ is shown in Fig.\ \ref{fig:rhoT} for one of the ensembles. 

The first correction $\delta\rho$ to $\rho_{0}$ is of order unity in $N_{1},N_{2}$, given by
\begin{align}
\delta\rho(T)={}&\tfrac{1}{4}(1-2/\beta)[\delta(1-T)-\delta(T-T_{c})]\nonumber\\
&-\tfrac{1}{2}(\gamma/\beta)\delta(1-T)\nonumber\\
&+\frac{1}{2\pi}(\gamma/\beta)\frac{\Theta(1-T)\Theta(T-T_{c})}{\sqrt{(1-T)(T-T_{c})}}.\label{deltarho}
\end{align}
We will use this expression in Sec.\ \ref{largenumber} to calculate the weak localization effect on the thermal conductance.

\section{Distribution of the thermal conductance}
\label{secPG}

\subsection{Minimal channel number}
\label{minimal}

The strikingly different probability distributions \eqref{PTDyson} and \eqref{PTAZ} in the normal and superconducting ensembles apply to transmission between contacts with a single (possibly degenerate) non-vanishing transmission eigenvalue. For the normal ensembles a narrow point contact suffices to reach this single-channel limit. In the superconducting ensembles a narrow point contact is not in general sufficient, because electrons and holes may still contribute independently to the thermal conductance. 

Consider the Andreev quantum dot of Fig.\ \ref{fig:AQD}. The minimal number of propagating modes incident on the quantum dot from each of the two leads is $2\times 2=4$: a factor-of-two counts the spin directions, and another factor-of-two the electron-hole degrees of freedom. In the CQE and T-CQE the four transmission eigenvalues are all degenerate, so we have reached the single-channel limit where the distribution \eqref{PTAZ} applies.

The situation is different in the CRE and T-CRE. In the T-CRE two of the four transmission eigenvalues are independent (and a two-fold Kramers degeneracy remains). In the CRE all four transmission eigenvalues are independent, but two of the four can be eliminated by spin-polarizing the leads by means of a sufficiently strong magnetic field. So the case with two independent transmission eigenvalues (with degeneracy factor $d=2$ for the T-CRE) is minimal in the Andreev quantum dot with broken spin-rotation symmetry. 

\begin{figure}[tb]
\centerline{\includegraphics[width=0.9\linewidth]{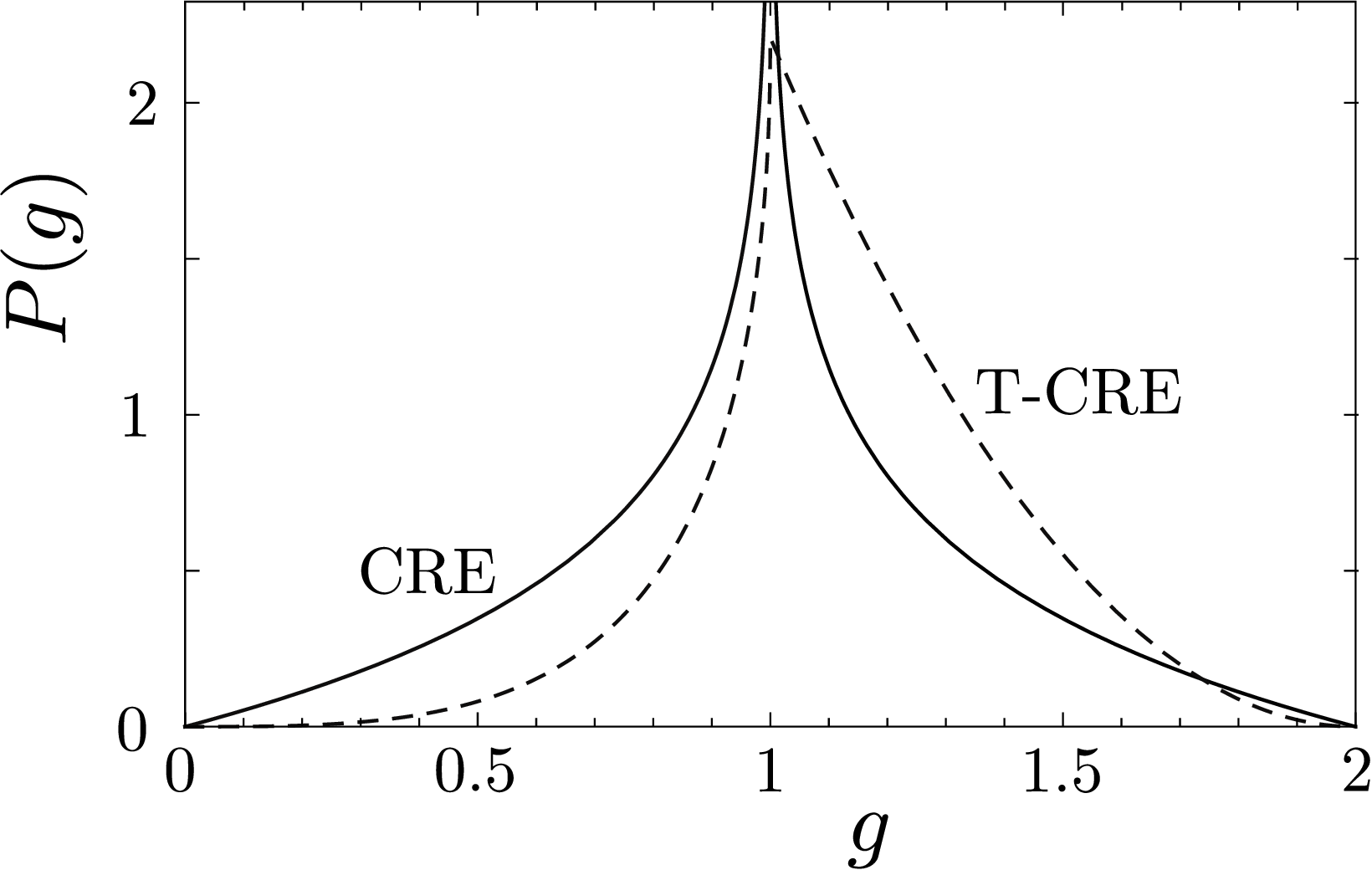}}
\caption{\label{fig:rhoPlot2}
Probability distribution of the dimensionless thermal conductance in the two ensembles with broken spin-rotation symmetry, for two independent transmission eigenvalues ($N_{1}=N_{2}=2$). This is the minimal channel number in an Andreev quantum dot. To reach the single-channel case in the CRE or T-CRE ($N_{1}=N_{2}=1$, plotted in Fig.\ \ref{fig:pdfPlot}) one needs a topological phase of matter, as discussed in Sec.\ \ref{sec:topo}. } 
\end{figure}

We have calculated the corresponding probability distribution of the (dimensionless) thermal conductance $g=T_{1}+T_{2}$ by integrating over the transmission eigenvalue distribution \eqref{pdf}. The result, plotted in Fig.\ \ref{fig:rhoPlot2}, has a singularity at $g=1$, in the form of a divergence in the CRE and a cusp in the T-CRE. It is entirely different from the distribution in the single-channel case (see Fig.\ \ref{fig:pdfPlot}). How to reach the single-channel limit in the CRE and T-CRE using topological phases of matter is described in Sec.\ \ref{sec:topo}.

\subsection{Large number of channels}
\label{largenumber}

In the limit $N_{1},N_{2}\gg 1$ of a large number of channels the distribution of the thermal conductance is a narrow Gaussian. We consider first the average and then the variance of this distribution.

The average conductance can be calculated by integrating over the eigenvalue density $\rho(T)$ of Sec.\ \ref{sec_rhoT}. We write the average of the dimensionless thermal conductance $g=G/dG_{0}$ as $\langle g\rangle=g_{0}+\delta g$, where $g_{0}$ is the leading order term for large $N_{1},N_{2}$ and $\delta g$ is the first correction. From Eqs.\ \eqref{rho0}--\eqref{deltarho} we obtain
\begin{align}
g_{0}={}&\frac{N_{1}N_{2}}{N_{1}+N_{2}},\label{g0result}\\
\delta g ={}&\frac{1}{\beta} (\beta-2-\gamma)\frac{N_{1}N_{2}}{(N_{1}+N_{2})^{2}}.\label{deltagresult}
\end{align}
The result \eqref{deltagresult} for $\delta g$ in the zero-dimensional regime of a quantum dot has the same dependence on the symmetry indices as in the one-dimensional wire geometry studied by Brouwer et al.\cite{Bro00}

Filling in the values of $\beta$, $\gamma$, and $d$ in the four superconducting ensembles from Table \ref{tab:table2}, we see that (for $N_{1}=N_{2}$)
\begin{equation}
\delta G = \begin{cases}
0 & \mbox{in the CRE and CQE},\\
-G_{0}/2 & \mbox{in the T-CQE},\\
G_{0}/4 & \mbox{in the T-CRE}.
\end{cases}\label{deltaGCE}
\end{equation}
This is fully analogous to the weak (anti)localization effect for the electrical conductance (with $G_{0}=e^{2}/h$) in the non-superconducting ensembles.\cite{Bee97} Without time-reversal symmetry (in the CRE, CQE, and CUE) there is no effect ($\delta G=0$), with both time-reversal and spin-rotation symmetry (in the T-CQE and COE) there is weak localization ($\delta G<0$) and with time-reversal symmetry but no spin-rotation symmetry (in the T-CRE and CSE) there is weak antilocalization ($\delta G>0$).

Turning now to the variance, we address the thermal analogue of universal conductance fluctuations. It is a central result of RMT\cite{Bee97} that the Gaussian distribution of $g$ has a variance of order unity in the large $N$-limit, determined entirely by the eigenvalue repulsion factor $\prod_{i<j}|T_{i}-T_{j}|^{\beta}$ in the probability distribution \eqref{pdf}. The $\gamma$-dependent factors plays no role. The result of the Dyson ensembles,\cite{Bar94,Jal94}
\begin{equation}
{\rm Var}\,g=\frac{2(N_{1}N_{2})^{2}}{\beta(N_{1}+N_{2})^{4}},\label{varg}
\end{equation}
therefore still applies in the Altland-Zirnbauer ensembles. 

For $N_{1}=N_{2}$ we find the variance of the thermal conductance ${\rm Var}\,G=G^2_{0}/p$ with $p=8,4,2,1$ in, respectively, the CRE, T-CRE, CQE, T-CQE. Breaking of time-reversal symmetry thus reduces the variance of the thermal conductance in the superconducting ensembles by a factor of two, while breaking of spin-rotation symmetry reduces it by a factor of four. This is fully analogous to the electrical conductance in the non-superconducting ensembles.

\subsection{Arbitrary number of channels}
\label{arbitraryN}

While the results from the previous subsection for the average and variance of the thermal conductance hold in the limit of a large number of channels, it is also possible to derive exact results for arbitrary $N_1,N_2$. Following the method described in Ref.\ \onlinecite{Sav06}, the moments of $g$ can be evaluated using the Selberg integral.\cite{Meh91} We find 
\begin{align}
\langle g \rangle ={}&  \frac{N_1 N_2}{N_{\rm t}+\xi} ,
\label{gSelberg}\\
{\rm Var}\, g  
={}&\frac{2N_{1}N_{2}  (N_{1}+\xi)(N_{2}+\xi)}{\beta(N_{\rm t}-1+\xi)(N_{\rm t}+\xi)^2(N_{\rm t}+\xi+2/\beta)},
\label{vargSelberg}
\end{align}
where we abbreviated $N_{t}=N_{1}+N_{2}$ and $\xi=(2-\beta+\gamma)/\beta$. One readily checks that the large-$N$ limits \eqref{g0result}, \eqref{deltagresult}, and \eqref{varg} are consistent with Eqs.\ \eqref{gSelberg} and \eqref{vargSelberg}. 

\section{How to reach the single-channel limit using topological phases}
\label{sec:topo}

\begin{figure}[tb]
\includegraphics[width=0.8\linewidth]{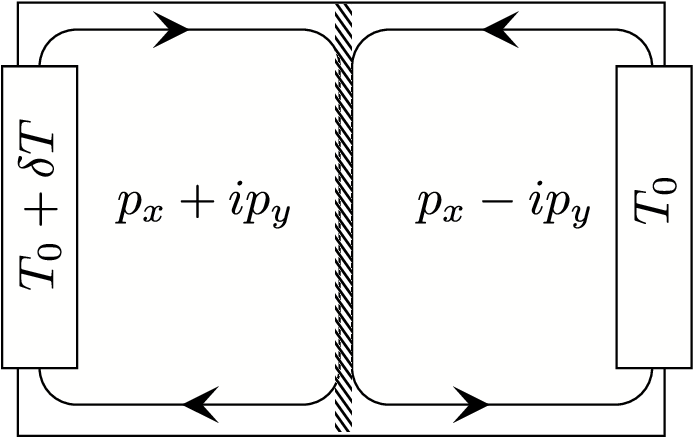}
\caption{\label{fig:topoCRE}
Realization of single-channel transmission in the CRE, following Ref.\ \onlinecite{Ser10}. The arrows indicate the direction of propagation of chiral Majorana modes at the edges of a $p_{x}\pm ip_{y}$-wave superconductor. The shaded strip at the center represents a disordered boundary between two domains of opposite chirality. The thermal conductance is measured between two reservoirs at a temperature difference $\delta T$, and has the single-channel distribution \eqref{PTAZ} (with $\beta=1$, $\gamma=-1$).
}
\end{figure}

As explained in Sec.\ \ref{minimal}, the single-channel distribution \eqref{PTAZ} of the thermal conductance can only be realized in an Andreev quantum dot in two of the four superconducting ensembles: CQE and T-CQE. The minimal channel number in the CRE and T-CRE is two, with an entirely different conductance distribution (compare Figs.\ \ref{fig:pdfPlot} and \ref{fig:rhoPlot2}). Here we show how this fermion doubling can be avoided using topological insulators or superconductors.

Consider first the CRE. To have just a single nonzero transmission eigenvalue we need incoming and outgoing modes that contain only half the degrees of freedom of spin-polarized electrons. These so-called Majorana modes propagate along the edge of a two-dimensional spin-polarized-triplet, $p_{x}\pm ip_{y}$-wave superconductor.\cite{Vol03,Kal09} Following Ref.\ \onlinecite{Ser10}, we consider the scattering geometry shown in Fig.\ \ref{fig:topoCRE}. The role of the quantum dot is played by a disordered domain wall between \textit{p}-wave superconductors of opposite chirality. The system has two incoming and two outgoing Majorana modes, with a $2\times 2$ scattering matrix in the CRE. The thermal conductance between the two domains has the single-channel distribution \eqref{PTAZ} (with $\beta=1$, $\gamma=-1$).  

\begin{figure}[tb]
\centerline{\includegraphics[width=0.8\linewidth]{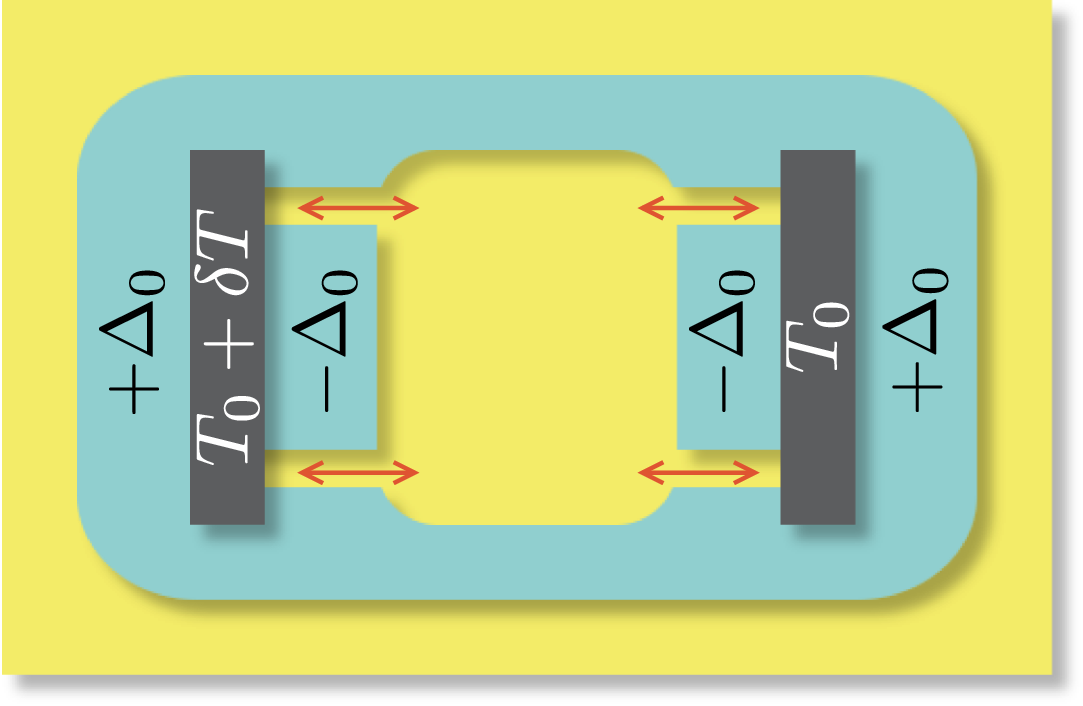}}
\caption{\label{fig:topoTCRE}
Realization of single-channel transmission in the T-CRE. The conducting surface of a topological insulator is partially covered by an \textit{s}-wave superconductor, with order parameter $\pm\Delta_{0}$. Two contacts at temperature difference $\delta T$ inject quasiparticles via two pairs of helical Majorana modes (indicated by arrows). For chaotic scattering in the central region, the thermal conductance is given by the single-channel distribution \eqref{PTAZ} (with $\beta=2$, $\gamma=-1$).
}
\end{figure}

We now turn to the T-CRE. For a single two-fold degenerate transmission eigenvalue we need a $4\times 4$ scattering matrix. Time-reversal invariant scattering in this single-channel limit can be achieved if one uses \textit{helical} Majorana modes (propagating in both directions) instead of \textit{chiral} Majorana modes (propagating in a single direction only). These can be realized using \textit{s}-wave superconductors deposited on the two-dimensional conducting surface of a three-dimensional topological insulator.\cite{Fu08} 

The scattering geometry is illustrated in Fig.\ \ref{fig:topoTCRE}. The helical Majorana modes propagate along a channel with superconducting boundaries having a phase difference of $\pi$ (order parameter $\pm\Delta_{0}$). Two normal-metal contacts at a temperature difference $\delta T$ inject quasiparticles via a pair of these modes into a region with chaotic scattering (provided by irregularly shaped boundaries or by disorder). The $\pi$ phase difference of the superconductors that form the boundaries of the quantum dot also ensures that there is no excitation gap in that region. There are four incoming and four outgoing Majorana modes, so the scattering matrix has dimension $4\times 4$ and the thermal conductance has the single-channel T-CRE distribution \eqref{PTAZ} (with $\beta=2$, $\gamma=-1$).

The geometry of Fig.\ \ref{fig:topoTCRE} also provides an alternative way to reach the single-channel limit in the CRE. One then needs to replace the two superconducting islands having order parameter $-\Delta_{0}$ by ferromagnetic insulators. The Majorana modes transform from helical to chiral\cite{Fu08} and one has essentially the same scattering geometry as in Fig.\ \ref{fig:topoCRE} --- but with \textit{s}-wave rather than \textit{p}-wave superconductors. 

\section{Conclusion}
\label{conclusion}

In conclusion, we have obtained the distribution of transmission eigenvalues for low-energy chaotic scattering in the four superconducting ensembles. From this distribution all moments of the thermal conductance of an Andreev quantum dot can be calculated. In the limit of a large number of scattering channels the phenomena of weak (anti)-localization and mesoscopic fluctuations are analogous to those for the electrical conductance in the non-superconducting ensembles. The opposite single-channel limit, however, shows striking differences. Most notably, in the absence of time-reversal symmetry, the thermal conductance distribution is either peaked or suppressed at minimal and maximal conductance, while the corresponding distribution of the electrical conductance is completely uniform.

While Andreev quantum dots with multiple scattering channels can be realized in a two-dimensional electron gas with \textit{s}-wave superconductors, the single-channel limit is out of reach in these systems in the absence of spin-rotation symmetry because of a fermion doubling problem. We have shown how Majorana modes at the interface between different topological phases can be used to overcome this problem.

In closing we point to the possibility to realize the four superconducting ensembles in graphene, where a strong proximity effect to \textit{s}-wave superconductors has been demonstrated.\cite{Hee07} An Andreev quantum dot in graphene could be created using superconducting boundaries,\cite{Cse09} as in Fig.\ \ref{fig:topoTCRE}. Since spin-orbit coupling is ineffective in graphene, only the two ensembles which preserve spin-rotation symmetry (CQE and T-CQE) are accessible in principle. However, if intervalley scattering is sufficiently weak (on the time scale set by the dwell time in the quantum dot), then the sublattice degree of freedom can play the role of the electron spin. This pseudospin is strongly coupled to the orbit, so one can then access the two ensembles with broken spin-rotation symmetry (CRE and T-CRE).

It is an interesting question to ask whether the single-channel limit might be reachable in graphene. For the CQE and T-CQE we need strong intervalley scattering, to remove the valley degeneracy. For the T-CRE we need weak intervalley scattering, and could use the very same setup as in Fig.\ \ref{fig:topoTCRE}. One can then do without a topological insulator, because the helical Majorana modes exist also in graphene at the interface between superconductors with a $\pi$ phase difference.\cite{Tit07} For the CRE, however, weak intervalley scattering is not enough. We would also need to convert the helical Majorana mode into a chiral mode, which we do not know how to achieve without a topological phase.

\acknowledgments

We thank A. R. Akhmerov for valuable discussions. This research was supported by the Dutch Science Foundation NWO/FOM and by an ERC Advanced Investigator Grant.

\appendix

\section{Calculation of the transmission eigenvalue distribution}
\label{calculationPT}

We briefly outline how to obtain the distribution \eqref{pdf} of the transmission eigenvalues from the invariant measure. (For a more detailed presentation of this type of calculation we refer to a textbook.\cite{For10}) One goes through the following steps. The polar decomposition of $S$ provides us with a parametrization in terms of the transmission eigenvalues $T_{i}$ and angular parameters $p_{i}$. We express the invariant measure $d\mu(S)$ in terms of these parameters via the metric tensor $m$: $d\mu(S)=\sqrt{{\rm det}\,m} \prod_{i}dx_{i}$, where $\{x_{i}\}$ denotes the full set of parameters $\{T_{i},p_{i}\}$ and $m$ is defined by ${\rm Tr}\,(dS^{\dag}dS)=\sum_{ij}m_{ij}dx_{i} dx_{j}$. Upon integration over the $p_{i}$'s we obtain the required distribution $P(\{T_{i}\})$.  

Starting from the first step, the polar decomposition reads
\begin{equation}
S=
\begin{pmatrix}
U_{1} &0\\
0 & U_{2}
\end{pmatrix}
\begin{pmatrix}
\sqrt{1-\Lambda \Lambda^{T}} & i \Lambda\\
i \Lambda^{T} &\sqrt{1-\Lambda^{T}\Lambda}
\end{pmatrix}
\begin{pmatrix}
V_{1}^{\dag} & 0\\
0 & V_{2}^{\dag}
\end{pmatrix},
\label{PolarDec}
\end{equation}
where the $N_{1}\times N_{2}$ matrix $\Lambda$ has elements $\Lambda_{jk}=\sqrt{T_{j}}\delta_{jk}$. Referring to Table \ref{tab:table2}, the transmission eigenvalues have a twofold electron-hole degeneracy in classes C and CI, as a direct consequence of the fact that the matrix elements can be represented by (real) quaternions. In addition, there is a twofold spin degeneracy because spin-rotation symmetry is preserved. In class DIII, the presence of time-reversal symmetry produces a twofold Kramers degeneracy of the transmission eigenvalues.
(We focus on the situation where $N_1$ and $N_2$ are even.)
 The unitary matrices $U_{n}$ and $V_{n}$ are orthogonal in classes D and DIII and symplectic in classes C and CI. They are independent in classes D and C. In class DIII one has $V_{n}^\dag=\sigma_{2}U_{n}^{T}\sigma_{2}$, while in class CI $V_{n}^{\dag}=U_{n}^{*}$.

The following steps are straightforward, apart from one complication. In the polar decomposition, the set of $T_{i}$'s and the matrices $U_{n}$ and $V_{n}$ introduce more parameters than the number of independent degrees  of freedom of the scattering matrix. The metric tensor, however, is defined through the derivatives of $S$ with respect to the set of its independent parameters. Keeping $\{T_{i}\}$ in our parametrization, we define the angular parameters $\{p_{i}\}$ as independent combinations of the matrix elements of $\delta U_{n}=U_{n}^{\dag} dU_{n}$ and $\delta V_{n}=V_{n}^{\dag} dV_{n}$. In this way, the subsequent integration over these degrees of freedom does not involve dependencies on the $T_i$'s. The integration over these parameters thus only produces an irrelevant normalization constant and need not be carried out explicitly.


\begin{thebibliography}{99}
\bibitem{note1} The basis in which $S$ is real is constructed by taking the combinations $|e\rangle+|h\rangle$, $i|e\rangle-i|h\rangle$ of the electron and hole states $|e\rangle$, $|h\rangle$ with the same spin and mode quantum numbers. The corresponding creation and annihilation operators are identical, so these basis states represent Majorana fermions.
\end{thebibliography}
\end{document}